\documentstyle[twocolumn,pra,aps,psfig]{revtex}

\begin{document}
\title{Effective mass in quantum effects of radiation pressure}
\author{M.\ Pinard, Y.\ Hadjar and A.\ Heidmann\thanks{%
e-mail : pinard, hadjar or heidmann@spectro.jussieu.fr}}
\address{Laboratoire Kastler Brossel\thanks{%
Laboratoire de l'Universit\'{e} Pierre et Marie Curie et de l'Ecole Normale
Sup\'{e}rieure associ\'{e} au Centre National de la Recherche Scientifique},
Case 74, 4 place Jussieu, F75252\ Paris Cedex 05, France}
\maketitle

\begin{abstract}
We study the quantum effects of radiation pressure in a high-finesse cavity
with a mirror coated on a mechanical resonator. We show that the
optomechanical coupling can be described by an effective susceptibility
which takes into account every acoustic modes of the resonator and their
coupling to the light.\ At low frequency this effective response is similar
to a harmonic response with an effective mass smaller than the total mass of
the mirror.\ For a plano-convex resonator the effective mass is related to
the light spot size and becomes very small for small optical waists, thus
enhancing the quantum effects of optomechanical coupling.
\end{abstract}

\bigskip

Radiation pressure exerted by light plays an important role in quantum
limits of very precise optical measurements.\ Quantum noise in
interferometers has two fundamental sources, the photon noise of the laser
beam and the fluctuations of the mirror's position due to radiation
pressure.\ Both lead to a quantum limit for measurement sensitivity and
potential applications of squeezed states have motivated a large number of
works in quantum optics\cite{Caves81,Jaekel90}.

When a movable mirror is exposed to a laser beam, its position is coupled to
the laser intensity via radiation pressure\cite{Braginsky74}.\ This {\it %
optomechanical coupling} may be enhanced using a high-finesse optical cavity
that is very sensitive to small mirror displacements. Such a device is
equivalent to cavities containing a Kerr medium and it has been studied for
squeezing generation\cite{Heidmann94,Fabre94,Mancini97,Bose97} or quantum
nondemolition measurements\cite{Tombesi94,Pinard95,Heidmann97}.

Theoretical calculations of these effects are usually based on a description
of the mechanical motion as a single harmonic oscillator. This model is well
adapted to the motion of the center of mass for suspended mirrors, such as
the ones of gravitational wave antennas\cite{Bradaschia90,Abramovici92}.
Radiation pressure can however excite internal acoustic modes of the mirror
for which a simple description as a harmonic oscillator is not appropriate.
These internal vibrations induce a deformation of the mirror which can be
coupled to the light.

In this paper, we study the effect of optomechanical coupling on quantum
fluctuations when internal acoustic modes of the mirror are considered. We
show that the coupling can be described by an effective response of the
mirror which takes into account all acoustic modes and their spatial
matching with the light.\ Similar results have been obtained for the
Brownian motion of mirrors which have been studied for gravitational-wave
interferometers\cite{Bondu95,Gillepsie95,Levin98,Bondu98}.\ We show that the
same effective response describes both thermal and radiation pressure
effects.\ This response can be approximated to a harmonic response at low
frequencies and the effective mass of this equivalent pendulum can be much
smaller than the total mass of the mirror.\ This small effective mass
enhances the quantum effects of optomechanical coupling.

\begin{figure}
\centerline{\psfig{figure=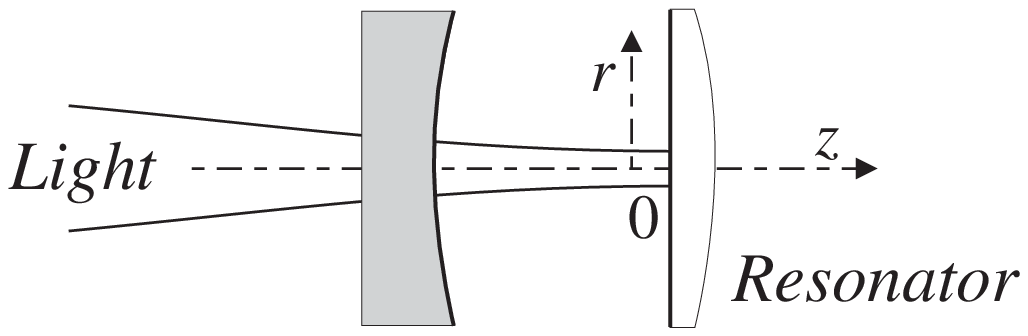,width=7cm}}
\caption{Model of a Fabry-Perot cavity with a movable mirror. The mirror is
coated on the plane side of a plano-convex mechanical resonator}
\label{Fig_Model}
\end{figure}

The system studied in this paper is sketched in figure \ref{Fig_Model}. It
consists of a single-port cavity with a movable mirror coated on the plane
side of a mechanical resonator. The geometry of the resonator determines the
spatial structure of the acoustic modes.\ The model presented in this paper
is however valid for any geometry. We first recall the basic properties of
the optomechanical coupling using a simple description in which both the
field and the mechanical resonator are treated as one-dimensional objects
and the mirror motion is described as a harmonic oscillator (Section \ref
{Section1}).

We then study the effect of internal acoustic modes by taking into account
the spatial structure of both the light and the resonator. We first
determine the effect of a deformation of the resonator on the light field
(Section \ref{Section2}).\ We then study the mechanical motion of the
resonator when its plane side is submitted to the radiation pressure\ of the
intracavity field. We finally define an effective susceptibility which
describes the optomechanical coupling between the light beam and the
resonator (Section \ref{Section3}). To illustrate the quantum effects of
radiation pressure we study the quantum noise reduction of the field
reflected by the cavity (Section \ref{Section4}).

We apply these results in the last section to the case of a plano-convex
resonator (Section \ref{Section5}). Acoustic modes are then confined near
the central axis of the resonator and their spatial structure can be
described by analytical expressions which are quite similar to gaussian
optical modes of a Fabry-Perot cavity.\ We show that this geometry leads to
a drastic reduction of the effective mass.

\section{One-dimensional model of optomechanical coupling}

\label{Section1}We recall in this section the basic properties of the
optomechanical coupling.\ For this purpose we neglect the internal motion of
the resonator and we assume that the mirror moves without any deformation of
its surface.\ The light is only sensitive to the mirror motion in the $z$
direction (Fig.\ \ref{Fig_Model}) and this motion can be completely
characterized by the position $z\left( t\right) $ of the mirror at time $t$.

The effect of the mirror motion on the intracavity field is a phase shift $%
\psi $ related to the change of the optical path followed by the light 
\begin{equation}
\psi \left( t\right) =2kz\left( t\right) \text{,}  \label{Eq_Psi}
\end{equation}
where $k$ is the wavevector of the light.\ The cavity detuning thus depends
on the cavity length and couples the field to the mirror motion.

For small mirror displacements, the linear response theory shows that the
Fourier transform $z\left[ \Omega \right] $ of the mirror position is
proportional to the applied force\cite{Landau58} 
\begin{equation}
z\left[ \Omega \right] =\chi \left[ \Omega \right] \left( 2\hbar kI\left[
\Omega \right] +F_{T}\left[ \Omega \right] \right) \text{,}  \label{Eq_z}
\end{equation}
where $\chi $ is the mechanical susceptibility of the mirror.\ If we assume
that the mechanical motion is harmonic, this susceptibility has a Lorentzian
shape 
\begin{equation}
\chi \left[ \Omega \right] =\frac{1}{M\left( \Omega _{M}^{2}-\Omega
^{2}-i\Omega _{M}^{2}\Phi \left[ \Omega \right] \right) }\text{,}
\label{Eq_ChiOH}
\end{equation}
where $M$ is the mass of the mirror and $\Omega _{M}$ its resonance
frequency. The loss angle $\Phi \left[ \Omega \right] $\ characterizes the
damping of the motion and is related to the quality factor $Q$\ of the
resonance by $\Phi \left[ \Omega _{M}\right] =1/Q$.\ The first force in (\ref
{Eq_z}) represents the radiation pressure exerted by the intracavity field.\
It is proportional to the momentum exchange $2\hbar k$ during a photon
reflection and to the light intensity $I$ normalized as the number of
photons reflected on the mirror by unit time. The second force $F_{T}$ is a
Langevin force describing the coupling of the resonator with a thermal
bath.\ Its spectrum $S_{T}\left[ \Omega \right] $ is related to the
mechanical susceptibility by the fluctuation-dissipation theorem\cite
{Landau58} 
\begin{equation}
S_{T}\left[ \Omega \right] =-\frac{2k_{B}T}{\Omega }%
\mathop{\rm Im}%
\left( \frac{1}{\chi \left[ \Omega \right] }\right) \text{,}  \label{Eq_ST}
\end{equation}
where $T$ is the temperature and $k_{B}$ the Boltzmann constant.

Equation (\ref{Eq_z}) shows that the radiation pressure exerted by the light
couples the mirror motion to the light intensity.\ The mean effect of this
coupling is a mean displacement $\overline{z}$ obtained from equation (\ref
{Eq_z}) by a statistical average of $z\left[ \Omega =0\right] $.\ From
equation (\ref{Eq_Psi}) this leads to a non-linear phase shift for the mean
field in the cavity equal to 
\begin{equation}
\psi _{NL}=4\hbar k^{2}\chi \left[ 0\right] \overline{I}\text{,}
\label{Eq_PsiNL}
\end{equation}
where $\overline{I}$ is the mean intracavity intensity.\ This
intensity-dependent phase shift is equivalent to an optical Kerr effect.\
The cavity with a movable mirror is thus equivalent to cavities containing a
Kerr medium which have been studied for squeezing generation\cite
{Collett85,Shelby86,Reynaud89} or\ QND\ measurements\cite
{Alsing88,Grangier89}. The main differences with a pure Kerr medium are due
to the dynamics of the moving mirror characterized by the frequency
dependence of the susceptibility $\chi \left[ \Omega \right] $ and to the
presence of thermal noise.

The non-linear phase shift $\psi _{NL}$ is an important parameter to
determine the efficiency of the optomechanical coupling.\ Quantum effects
are significant if this phase shift is of the order of the cavity losses\cite
{Heidmann97}. In this case the displacement $\overline{z}$ induced by the
mean radiation pressure is of the order of the width $\lambda /{\cal F}$ of
the cavity resonance ($\lambda $ is the optical wavelength and ${\cal F}$ is
the cavity finesse). This condition actually corresponds to the observation
of bistability in the cavity.\ As usual in quantum optics, quantum effects
are important near the bistability turning points.\ This condition depends
on the optical characteristics of the system, such as the cavity finesse or
the light intensity. It also depends on the mechanical properties of the
resonator (eq. \ref{Eq_PsiNL}).\ In particular, quantum effects induced by
optomechanical coupling are inversely proportional to the mass of the
resonator.

\section{Light reflexion on a moving mirror}

\label{Section2}We have shown in the previous section that the
optomechanical coupling is based on two complementary effects.\ The first
one is the phase shift of light induced by the mirror motion.\ The second
effect is that the mirror moves in response to radiation pressure. We
examine in this section the first effect when internal modes of the
resonator are taken into account.

Light is sensitive only to longitudinal displacements of the mirror.\ A
longitudinal deformation of the plane side of the resonator can be described
by its displacement $u\left( r,z=0,t\right) $ in the $z$-direction at every
radial point $r$ of the surface (the origin of the cylindrical coordinates
is taken at the center of the mirror, see figure \ref{Fig_Model}).\ The
electric field in front of the moving mirror can be written as 
\begin{equation}
E\left( r,t\right) =v_{0}\left( r\right) \alpha \left( t\right) e^{-i\omega
_{0}t}\text{,}
\end{equation}
where $\alpha \left( t\right) $ is the slowly varying amplitude of the
field, $\omega _{0}$ the optical frequency and $v_{0}\left( r\right) $ the
spatial structure of the beam in the $z=0$ plane.\ In the paraxial
approximation, the optical modes of the cavity are gaussian modes with their
waist at position $z=0$\cite{Kogelnik66}. Assuming the incident beam matched
to the fundamental mode of the cavity, the spatial structure $v_{0}\left(
r\right) $ of the field is given by 
\begin{equation}
v_{0}\left( r\right) =\frac{\sqrt{2/\pi }}{w_{0}}e^{-r^{2}/w_{0}^{2}}\text{,}
\label{Eq_v0}
\end{equation}
where $w_{0}$ is the optical waist which depends on the geometry of the
cavity.

The spatial structure of the reflected field is modified by the mirror
motion since the field wavefront is distorted and reproduces after
reflection the shape of the mirror.\ At every radial point $r$ of the
mirror, the optical path followed by the light is changed by the
displacement and\ the field experiences a local phase-shift equal to $%
2ku\left( r,z=0,t\right) .$ The reflected field $E^{\prime }\left(
r,t\right) $ is then equal to 
\begin{equation}
E^{\prime }\left( r,t\right) =v_{0}\left( r\right) \alpha \left( t\right)
e^{-i\omega _{0}t}e^{2iku\left( r,z=0,t\right) }\text{.}
\end{equation}
Noting $\left\{ v_{n}\left( r\right) \right\} $ the basis of gaussian modes
of the cavity, one can write the reflected field as a sum over all modes 
\begin{equation}
E^{\prime }\left( r,t\right) =\sum_{n}\left\langle v_{0}e^{2iku\left(
z=0,t\right) },v_{n}\right\rangle v_{n}\left( r\right) \alpha \left(
t\right) e^{-i\omega _{0}t}\text{,}  \label{Eq_Erefl}
\end{equation}
where the brackets stand for the overlap integral in the $z=0$ plane 
\begin{equation}
\left\langle f,g\right\rangle =\int_{z=0}d^{2}rf\left( r\right) g\left(
r\right) \text{.}  \label{Eq_Scalar}
\end{equation}
Equation (\ref{Eq_Erefl}) shows that the mirror deformation induces a
diffusion of the light into all optical modes. This diffusion is however
limited by the cavity and it becomes negligible for a non-degenerate and
high-finesse cavity. In this case, modes in the sum (\ref{Eq_Erefl}) evolve
at a frequency about the resonance frequency $\omega _{0}$ of the
fundamental mode whereas the differences between the resonance frequencies
of these modes are large compared to the cavity bandwidth. As a consequence,
all modes except the fundamental one are filtered by the cavity bandwidth
and cannot propagate in the cavity. One can show that the diffusion in those
modes is equivalent to losses for the fundamental mode and that these losses
become negligible for a high-finesse cavity\cite{Courty96}.\ Only the
fundamental mode has thus a significant contribution in the sum (\ref
{Eq_Erefl}) and we obtain 
\begin{equation}
E^{\prime }\left( r,t\right) =\left\langle v_{0}e^{2iku\left( z=0,t\right)
},v_{0}\right\rangle E\left( r,t\right) \text{.}
\end{equation}
For small displacements $u$\ this expression can be approximated to 
\begin{equation}
E^{\prime }\left( r,t\right) \approx \left[ 1+2ik\left\langle u\left(
z=0,t\right) ,v_{0}^{2}\right\rangle \right] E\left( r,t\right) \text{.}
\end{equation}
The field thus experiences a global phase-shift $\psi \left( t\right) $
which can be written as 
\begin{equation}
\psi \left( t\right) =2k\widehat{u}\left( t\right) \text{,}  \label{Eq_Psi_u}
\end{equation}
where $\widehat{u}\left( t\right) $ is the displacement of the mirror
averaged over the optical waist 
\begin{equation}
\widehat{u}\left( t\right) =\left\langle u\left( z=0,t\right)
,v_{0}^{2}\right\rangle \text{.}  \label{Eq_u}
\end{equation}
The effect of the mirror motion on the intracavity field is thus equivalent
to the one obtained in the one-dimensional model. It corresponds to a phase
shift given by an equation similar to (\ref{Eq_Psi}) where the
one-dimensional displacement $z\left( t\right) $ is replaced by the averaged
displacement $\widehat{u}\left( t\right) $. All results concerning the
effect of the mirror motion on the field can thus be generalized to a
mechanical resonator.\ The light is only sensitive to the displacement $%
\widehat{u}$ of the mirror which takes into account the spatial overlap
between the intracavity field and the mirror motion.\ In the next section we
determine this displacement when the resonator is submitted to the radiation
pressure of the intracavity field.

\section{Radiation pressure effects}

\label{Section3}Spatial and frequency characteristics of the acoustic modes
depend on the geometry and acoustic properties of the mechanical resonator.
To determine the effect of radiation pressure on the mirror motion, it is
only necessary to assume that these modes are described by a set of
displacements $\left\{ \overrightarrow{u}_{n}\left( \overrightarrow{r}%
\right) \right\} $ which forms a basis of the resonator motion.\ Each mode $%
n $ obeys a propagation equation inside the resonator\cite{Landau58b} 
\begin{equation}
c_{t}^{2}\Delta \overrightarrow{u}_{n}+\left( c_{l}^{2}-c_{t}^{2}\right) 
\overrightarrow{\nabla }\left( \overrightarrow{\nabla }.\overrightarrow{u}%
_{n}\right) +\Omega _{n}^{2}\overrightarrow{u}_{n}=0\text{,}
\label{Eq_Propag}
\end{equation}
and fulfills the boundary conditions corresponding to a free resonator (see
Appendix). In this equation $\Omega _{n}$ is the eigenfrequency of mode $n$\
and $c_{l}$, $c_{t}$ are the longitudinal and transverse sound velocities
related to the Lam\'{e} constants $\lambda $ and $\mu $ of the resonator and
to its density $\rho $ by 
\begin{mathletters}
\label{Eqs_Velocity}
\begin{eqnarray}
c_{l} &=&\sqrt{\left( \lambda +2\mu \right) /\rho }\text{,} \\
c_{t} &=&\sqrt{\mu /\rho }\text{.}
\end{eqnarray}

Any displacement $\overrightarrow{u}\left( \overrightarrow{r},t\right) $ can
be expressed as a linear combination of the acoustic modes $\overrightarrow{u%
}_{n}\left( \overrightarrow{r}\right) $ 
\end{mathletters}
\begin{equation}
\overrightarrow{u}\left( \overrightarrow{r},t\right) =\sum_{n}a_{n}\left(
t\right) \overrightarrow{u}_{n}\left( \overrightarrow{r}\right) \text{,}
\label{Eq_Decomp}
\end{equation}
where $a_{n}\left( t\right) $\ is the time-dependent amplitude of mode $n$.
Using this decomposition we can determine the evolution of $\overrightarrow{u%
}\left( \overrightarrow{r},t\right) $ when a radiation pressure force is
applied on the plane side of the resonator.\ The total energy $E$\ of the
resonator is the sum of the kinetic energy, the potential energy and the
energy associated with the external force.\ It can be decomposed in the
following form (see Appendix) 
\begin{eqnarray}
E &=&\sum_{n}\left\{ \frac{1}{2}M_{n}\left( \frac{da_{n}}{dt}\right) ^{2}+%
\frac{1}{2}M_{n}\Omega _{n}^{2}a_{n}^{2}\right.  \nonumber \\
&&\left. -\left\langle \overrightarrow{F}_{rad},\overrightarrow{u}%
_{n}\right\rangle a_{n}\right\} \text{.}  \label{Eq_Energy}
\end{eqnarray}
$M_{n}$\ represents the mass of the acoustic mode $n$\ which is proportional
to the volume of the mode inside the resonator 
\begin{equation}
M_{n}=\rho \int_{V}d^{3}r\left| \overrightarrow{u_{n}}\left( \overrightarrow{%
r}\right) \right| ^{2}\text{.}  \label{Eq_Mn}
\end{equation}
The term $\left\langle \overrightarrow{F}_{rad},\overrightarrow{u}%
_{n}\right\rangle $ in equation (\ref{Eq_Energy}) represents the spatial
overlap of the scalar product between the radiation pressure $%
\overrightarrow{F}_{rad}$\ and the acoustic mode $\overrightarrow{u}_{n}$\
(eq. \ref{Eq_Scalar}). The radiation pressure is directed along the $z$-axis
and its amplitude $F_{rad}\left( r,t\right) $\ at radial position $r$\ of
the plane side and at time $t$\ is related to the intracavity intensity $%
I\left( t\right) =\left| \alpha \left( t\right) \right| ^{2}$\ and to the
spatial structure $v_{0}$\ of the field by 
\begin{equation}
F_{rad}\left( r,t\right) =2\hbar kI\left( t\right) v_{0}^{2}\left( r\right) 
\text{.}  \label{Eq_Frad}
\end{equation}
The total energy\ (eq. \ref{Eq_Energy}) appears as the sum over all modes $n$%
\ of the energies of forced harmonic oscillators. From Hamilton's equations
one deduces that each mode amplitude $a_{n}\left( t\right) $\ obeys the
evolution equation 
\begin{equation}
\frac{d^{2}a_{n}}{dt^{2}}+\Omega _{n}^{2}a_{n}=\frac{1}{M_{n}}\left\langle 
\overrightarrow{F}_{rad},\overrightarrow{u}_{n}\right\rangle \text{.}
\label{Eq_Motion_an}
\end{equation}
This equation can be written in the Fourier space as 
\begin{equation}
a_{n}\left[ \Omega \right] =\chi _{n}\left[ \Omega \right] \left\langle 
\overrightarrow{F}_{rad}\left[ \Omega \right] ,\overrightarrow{u}%
_{n}\right\rangle \text{,}  \label{Eq_an1}
\end{equation}
where $\chi _{n}\left[ \Omega \right] =1/M_{n}\left( \Omega _{n}^{2}-\Omega
^{2}\right) $ is the susceptibility in absence of dissipation for a harmonic
oscillator of mass $M_{n}$ and eigenfrequency $\Omega _{n}$.\ The resonator
can thus be considered as a set of independent harmonic oscillators, each
oscillator being associated with an acoustic mode.\ These oscillators are
driven by an external force which corresponds to the projection of the
radiation pressure onto the spatial structure of the acoustic mode in the $%
z=0$\ plane.

Up to now we have assumed that the resonator has no damping. The coupling
with a thermal bath can be deduced from Navier-Stokes equation\cite
{Landau58b} or from a generalization of the approach used in the
one-dimensional model. Each acoustic mode is indeed equivalent to a harmonic
oscillator\ and its damping can be described by a dissipative part added to
the mechanical susceptibility and by an additional Langevin force. The
susceptibility $\chi _{n}$\ of the acoustic mode $n$\ has thus an expression
similar to the one-dimensional case 
\begin{equation}
\chi _{n}\left[ \Omega \right] =\frac{1}{M_{n}\left( \Omega _{n}^{2}-\Omega
^{2}-i\Omega _{n}^{2}\Phi _{n}\left[ \Omega \right] \right) }\text{,}
\label{Eq_Chin}
\end{equation}
where $\Phi _{n}\left[ \Omega \right] $\ is the loss angle of mode $n$. The
amplitude $a_{n}$\ of mode $n$\ is now given by 
\begin{equation}
a_{n}\left[ \Omega \right] =\chi _{n}\left[ \Omega \right] \left(
\left\langle \overrightarrow{F}_{rad}\left[ \Omega \right] ,\overrightarrow{u%
}_{n}\right\rangle +F_{T,n}\left[ \Omega \right] \right) \text{,}
\label{Eq_an2}
\end{equation}
where $F_{T,n}$\ is a Langevin force describing the coupling of mode $n$\
with the thermal bath. We assume that these forces are statistically
independent from each other and that their spectra are related to the
susceptibilities $\chi _{n}$ through fluctuations-dissipation theorem (eq. 
\ref{Eq_ST}). Acoustic modes are then independent and the equation of motion
for each amplitude $a_{n}$\ (eq.\ \ref{Eq_an2}) corresponds to the usual
expression for a damped harmonic oscillator driven by the projection of the
radiation pressure.

We have shown in the previous section that the effect of the mirror
deformation on the intracavity field only depends on the longitudinal
displacement $\widehat{u}\left( t\right) $ which corresponds to the
displacement in the $z=0$ plane averaged over the beam waist (eq. \ref{Eq_u}%
). From equations (\ref{Eq_u}), (\ref{Eq_Decomp}), (\ref{Eq_Frad}) and (\ref
{Eq_an2}) this displacement can be expressed in terms of the intracavity
intensity $I$ and of thermal fluctuations 
\begin{equation}
\widehat{u}\left[ \Omega \right] =\chi _{eff}\left[ \Omega \right] \left(
2\hbar kI\left[ \Omega \right] +F_{T}\left[ \Omega \right] \right) \text{.}
\label{Eq_u2}
\end{equation}
$\chi _{eff}$ appears in this equation as an effective susceptibility given
by 
\begin{equation}
\chi _{eff}\left[ \Omega \right] =\sum_{n}\left\langle
v_{0}^{2},u_{n}\right\rangle ^{2}\chi _{n}\left[ \Omega \right] \text{,}
\label{Eq_ChiEff}
\end{equation}
where $u_{n}$\ stands for the $z$-component of the acoustic mode $%
\overrightarrow{u}_{n}$. The effective susceptibility $\chi _{eff}$\ is then
equal to the sum of all susceptibilities $\chi _{n}$ weighted by the overlap
between the acoustic mode and the transverse intensity distribution $%
v_{0}^{2}$. The force $F_{T}$ in equation (\ref{Eq_u2}) is an effective
Langevin force related to the forces $F_{T,n}$ of each acoustic modes by 
\begin{equation}
F_{T}\left[ \Omega \right] =\sum_{n}\left\langle
v_{0}^{2},u_{n}\right\rangle \frac{\chi _{n}\left[ \Omega \right] }{\chi
_{eff}\left[ \Omega \right] }F_{T,n}\left[ \Omega \right] \text{.}
\label{Eq_FT}
\end{equation}
The spectrum of $F_{T}$\ can be determined using the independence of the
Langevin forces $F_{T,n}$ and their relation to susceptibilities $\chi _{n}$
(eq. \ref{Eq_ST}). We find that the force $F_{T}$ is related to $\chi _{eff}$
by the fluctuation-dissipation theorem 
\begin{equation}
S_{T}\left[ \Omega \right] =-\frac{2k_{B}T}{\Omega }%
\mathop{\rm Im}%
\left( \frac{1}{\chi _{eff}\left[ \Omega \right] }\right) \text{.}
\label{Eq_ST2}
\end{equation}
This means that the resonator in absence of external force is in
thermodynamic equilibrium at temperature $T$. Equation (\ref{Eq_u2}) also
shows that the effective susceptibility $\chi _{eff}$ describes both the
effects of radiation pressure and of thermal noise represented by the
Langevin force $F_{T}$.

Results obtained here are similar to the ones presented in section \ref
{Section1} for a one-dimensional model. The longitudinal displacement $%
\widehat{u}$\ is related to the intracavity intensity $I$\ and to thermal
fluctuations by an expression similar to equation (\ref{Eq_z}). The coupling
between the resonator and the gaussian light beam is completely described by
the effective susceptibility $\chi _{eff}$. This susceptibility indeed takes
into account all acoustic modes of the resonator and their spatial matching
with the light.\ Treatments made in the framework of the one-dimensional
model can thus be generalized by replacing the mirror displacement $z\left(
t\right) $\ by the longitudinal displacement $\widehat{u}\left( t\right) $\
and the harmonic susceptibility $\chi $\ by the effective susceptibility $%
\chi _{eff}$. This is illustrated in the next section where we study the
quantum noise reduction of the field reflected by the cavity.

\section{Quantum-noise reduction}

\label{Section4}Squeezed-state generation has already been studied for a
single-ended cavity containing a pure Kerr medium\cite
{Collett85,Shelby86,Reynaud89} or for a cavity with a harmonically suspended
mirror\cite{Fabre94}.\ In this section we extend these results to the case
of a mirror coated on a mechanical resonator. We thus consider the system
sketched in figure \ref{Fig_Model} and we determine the quantum fluctuations
of any quadrature of the reflected beam.

For a nearly resonant high-finesse cavity, we have the following relations
between the complex amplitudes $\alpha ^{in}$, $\alpha $ and $\alpha ^{out}$
of the incident, intracavity and reflected fields respectively 
\begin{mathletters}
\label{Eqs_Field}
\begin{eqnarray}
\tau \frac{d\alpha }{dt} &=&-\left[ \gamma -i\Psi \left( t\right) \right]
\alpha \left( t\right) +\sqrt{2\gamma }\alpha ^{in}\left( t\right) \text{,}
\label{Eq_da/dt} \\
\alpha ^{out}\left( t\right) &=&-\alpha ^{in}\left( t\right) +\sqrt{2\gamma }%
\alpha \left( t\right) \text{.}
\end{eqnarray}
The first relation determines the dynamics of\ the intracavity field $\alpha 
$. $\tau $ is the round trip time, $\gamma $ the damping rate of the cavity (%
$1-\gamma $ and $\sqrt{2\gamma }$ are respectively the reflection and
transmission of the input mirror, with $\gamma \ll 1$) and $\Psi $ is the
cavity detuning assumed small compared to $1$. $\Psi $ depends on the cavity
lentgh and couples the field to the mirror motion. It can be written as the
sum of\ the cavity detuning $\Psi _{0}$ without light and the effect of the
averaged displacement $\widehat{u}\left( t\right) $ (eq. \ref{Eq_Psi_u}) 
\end{mathletters}
\begin{equation}
\Psi =\Psi _{0}+2k\widehat{u}\text{.}
\end{equation}

The mean cavity detuning $\overline{\Psi }$ differs from $\Psi _{0}$ by the
contribution due to the mean mirror displacement.\ Using equation (\ref
{Eq_u2}) this contribution appears as a non-linear phase shift $\Psi _{NL}$\
related to the mean intracavity intensity $\overline{I}$\ by an expression
similar to equation (\ref{Eq_PsiNL}) 
\begin{mathletters}
\label{Eqs_Psi}
\begin{eqnarray}
\overline{\Psi } &=&\Psi _{0}+\Psi _{NL}\text{,} \\
\Psi _{NL} &=&4\hbar k^{2}\chi _{eff}\left[ 0\right] \overline{I}\text{.}
\label{Eq_PsiNL2}
\end{eqnarray}
The behavior of the mean fields is thus the same as the one for a cavity
containing a pure Kerr medium.\ In particular the non-linear phase shift is
responsible for the bistability of the system\cite{Fabre94}. From equations (%
\ref{Eqs_Field}) the mean intracavity intensity is related to the incident
power $P^{in}$ by 
\end{mathletters}
\begin{equation}
\left( \gamma ^{2}+\overline{\Psi }^{2}\right) \overline{I}=2\gamma \frac{%
\lambda }{hc}P^{in}\text{.}  \label{Eq_I}
\end{equation}
Due to the intensity dependence of the non-linear phase shift, one can
obtain different intracavity intensities for a given incident power.\ This
bistable behavior can be characterized by the inverse slope $\sigma
=dP^{in}/d\overline{I}$ of the curve giving the intracavity intensity as a
function of the incident power. $\sigma $ is proportional to 
\begin{equation}
\sigma \propto \gamma ^{2}+\overline{\Psi }^{2}+2\overline{\Psi }\Psi _{NL}%
\text{.}  \label{Eq_Sigma}
\end{equation}
The positive, negative, and null values of $\sigma $ correspond,
respectively, to stable branches, unstable branch, and turning points\cite
{Reynaud89}.

To determine\ the fluctuations $\delta \alpha ^{out}\left[ \Omega \right] $
of the field reflected by the cavity, we use the semiclassical method in
which quantum fluctuations are treated as classical random variables
associated with the Wigner distribution\cite{Reynaud89b,Reynaud92,Fabre92}.
Their evolution is deduced from the classical equations (\ref{Eqs_Field})
linearized around the mean state. We obtain 
\begin{mathletters}
\label{Eqs_Fluct}
\begin{eqnarray}
\left( \gamma -i\overline{\Psi }-i\Omega \tau \right) \delta \alpha \left[
\Omega \right] &=&\sqrt{2\gamma }\delta \alpha ^{in}\left[ \Omega \right] +i%
\overline{\alpha }\delta \Psi \left[ \Omega \right] \text{,} \\
\delta \alpha ^{out}\left[ \Omega \right] &=&\sqrt{2\gamma }\delta \alpha %
\left[ \Omega \right] -\delta \alpha ^{in}\left[ \Omega \right] \text{,} \\
\overline{\alpha }\delta \Psi \left[ \Omega \right] &=&\widetilde{\chi }%
_{eff}\left[ \Omega \right] \Psi _{NL}\left( \delta \alpha \left[ \Omega %
\right] +\delta \alpha ^{\ast }\left[ \Omega \right] \right)  \nonumber \\
&&+2k\overline{\alpha }\chi _{eff}\left[ \Omega \right] F_{T}\left[ \Omega %
\right] \text{,}
\end{eqnarray}
where $\overline{\alpha }$\ is the mean intracavity field and $\widetilde{%
\chi }_{eff}\left[ \Omega \right] =\chi _{eff}\left[ \Omega \right] /\chi
_{eff}\left[ 0\right] $ is the mechanical susceptibility normalized to 1 at
zero frequency. We deduce from these equations the input-output relations
for the field which give the output field fluctuations $\delta \alpha ^{out}$
as a function of the input field fluctuations $\delta \alpha ^{in}$ and of
the Langevin force $F_{T}$ 
\end{mathletters}
\begin{eqnarray}
\delta \alpha ^{out}\left[ \Omega \right] &=&\left\{ c_{1}\left[ \Omega %
\right] \delta \alpha ^{in}\left[ \Omega \right] +c_{2}\left[ \Omega \right]
\delta \alpha ^{in\ast }\left[ \Omega \right] \right.  \nonumber \\
&&\left. +c_{T}\left[ \Omega \right] F_{T}\left[ \Omega \right] \right\} 
\text{,}  \label{Eq_daout}
\end{eqnarray}
where the coefficients $c_{1}\left[ \Omega \right] $, $c_{2}\left[ \Omega %
\right] $ and $c_{T}\left[ \Omega \right] $ depend on the system parameters 
\begin{mathletters}
\label{Eqs_Coeffs}
\begin{eqnarray}
c_{1}\left[ \Omega \right] &=&\frac{1}{\Delta }\left\{ \left( \gamma +i%
\overline{\Psi }\right) \left( \gamma +i\overline{\Psi }+2i\Psi _{NL}%
\widetilde{\chi }_{eff}\left[ \Omega \right] \right) \right.  \nonumber \\
&&\left. +\left( \Omega \tau \right) ^{2}\right\} \text{,} \\
c_{2}\left[ \Omega \right] &=&\frac{2i}{\Delta }\gamma \Psi _{NL}\widetilde{%
\chi }_{eff}\left[ \Omega \right] \text{,} \\
c_{T}\left[ \Omega \right] &=&\frac{2i}{\Delta }\sqrt{2\gamma }k\overline{%
\alpha }\left( \gamma +i\overline{\Psi }-i\Omega \tau \right) \chi _{eff}%
\left[ \Omega \right] \text{,} \\
\Delta &=&\left( \gamma -i\Omega \tau \right) ^{2}+\overline{\Psi }^{2}+2%
\overline{\Psi }\Psi _{NL}\widetilde{\chi }_{eff}\left[ \Omega \right] \text{%
.}
\end{eqnarray}

From these relations one can determine the spectrum $S_{\theta }^{out}$ for
any quadrature $\alpha _{\theta }^{out}$ of the reflected field defined by 
\end{mathletters}
\begin{mathletters}
\label{Eqs_daout}
\begin{eqnarray}
\delta \alpha _{\theta }^{out}\left[ \Omega \right] &=&e^{-i\theta }\delta
\alpha ^{out}\left[ \Omega \right] +e^{i\theta }\delta \alpha ^{out\ast }%
\left[ \Omega \right] \text{,} \\
\overline{\delta \alpha _{\theta }^{out}\left[ \Omega \right] \delta \alpha
_{\theta }^{out}\left[ \Omega ^{\prime }\right] } &=&2\pi \delta \left(
\Omega +\Omega ^{\prime }\right) S_{\theta }^{out}\left[ \Omega \right] 
\text{,}
\end{eqnarray}
where the bar stands for the average over the Wigner distribution. Equations
(\ref{Eq_daout}) and (\ref{Eqs_Coeffs}) show that the quantum properties of
the reflected field only depend on a few parameters.\ The cavity is
characterized by the damping $\gamma $\ which is related to its finesse $%
{\cal F}=\pi /\gamma $, by the cavity bandwidth $\Omega _{cav}=\gamma /\tau $
and by the mean detuning $\overline{\Psi }$. The mirror motion is described
by the non-linear phase shift $\Psi _{NL}$\ and by the frequency dependence $%
\widetilde{\chi }_{eff}$\ of the mechanical response. The last coefficient $%
c_{T}$\ describes thermal effects associated with the Brownian motion of the
mirror. To obtain quantitative values for the noise reduction, it is
necessary to determine the mechanical response of the resonator which may
depend on its geometry.\ We study in the next section the case of a
plano-convex resonator.

Note finally that the differences with a pure Kerr effect are only due to
the frequency dependence of the mechanical response and to thermal
fluctuations.\ One gets the usual expressions for a Kerr medium by taking $%
\widetilde{\chi }_{eff}\left[ \Omega \right] =1$ and $F_{T}\left[ \Omega %
\right] =0$ in equations (\ref{Eq_daout}) and (\ref{Eqs_Coeffs}).

\section{Optomechanical coupling with a plano-convex resonator}

\label{Section5}We consider in this section that the resonator has a plano-convex geometry
with a mirror coated on its plane side (Fig. \ref{Fig_Model}). We first
determine the analytical expression of the effective susceptibility.\ We
then study the quantum-noise reduction of the field reflected by the cavity
and we finally derive the effective mass associated with the optomechanical
coupling.

\subsection{Effective susceptibility}

If the resonator thickness $h_{0}$ is much smaller than the curvature radius 
$R$ of the convex side, the propagation equation (\ref{Eq_Propag}) can be
solved using a paraxial approximation and one gets analytical expressions
for the acoustic modes corresponding to gaussian modes\cite{Wilson74}. Only
modes that have a non-zero overlap with the light intensity contribute to
the effective susceptibility (eq.\ \ref{Eq_ChiEff}). As a consequence we
disregard shear modes which induce no longitudinal displacement and we
consider only the compression modes that have a cylindrical symmetry. Those
modes are defined by two integers, a longitudinal index $n$ and a transverse
index $p$, and the longitudinal displacement $u_{n,p}(r,z)$ at point of
radial coordinate r and axial coordinate z is given by 
\end{mathletters}
\begin{equation}
u_{n,p}\left( r,z\right) =e^{-r^{2}/w_{n}^{2}}L_{p}\left(
2r^{2}/w_{n}^{2}\right) \cos \left( \frac{n\pi }{h\left( r\right) }z\right) 
\text{.}  \label{Eq_unp}
\end{equation}
$u_{n,p}$\ is composed of a transverse gaussian structure with a waist $%
w_{n} $, a transverse Laguerre polynomial $L_{p}$ and a cosine in the
propagation direction. $h\left( r\right) $ is the resonator thickness at
radial position r\ given by 
\begin{equation}
h\left( r\right) \approx h_{0}-\frac{r^{2}}{2R}\text{.}
\end{equation}
Acoustic waists $w_{n}$ depend on the longitudinal index $n$ 
\begin{equation}
w_{n}^{2}=\frac{2h_{0}}{n\pi }\sqrt{Rh_{0}}\text{.}
\end{equation}
Acoustic modes $u_{n,p}$ are solution of the propagation equation inside the
resonator and each mode evolves with an eigenfrequency $\Omega _{n,p}$ given
by 
\begin{equation}
\Omega _{n,p}^{2}=\Omega _{M}^{2}\left[ n^{2}+\frac{2}{\pi }\sqrt{\frac{h_{0}%
}{R}}n\left( 2p+1\right) \right] \text{,}  \label{Eq_Onp}
\end{equation}
where $\Omega _{M}=\pi c_{l}/h_{0}$.

One can now derive an analytical expression for the effective susceptibility
as an infinite sum over all modes $\left\{ n,p\right\} $\ (eq.\ \ref
{Eq_ChiEff}). In this expression the mass $M_{n}$ of the acoustic mode $%
\left\{ n,p\right\} $\ (eq.\ \ref{Eq_Mn}) only depends on the longitudinal
index $n$ and is equal to 
\begin{equation}
M_{n}=\frac{\pi }{4}\rho h_{0}w_{n}^{2}\text{.}  \label{Eq_Mnp}
\end{equation}
This mass is proportional to the volume of the acoustic mode and is smaller
than the total mass of the resonator. We can also derive the overlap between
acoustic and optical modes 
\begin{equation}
\left\langle v_{0}^{2},u_{n,p}\right\rangle =\frac{2w_{n}^{2}}{%
2w_{n}^{2}+w_{0}^{2}}\left( \frac{2w_{n}^{2}-w_{0}^{2}}{2w_{n}^{2}+w_{0}^{2}}%
\right) ^{p}\text{.}  \label{Eq_Overlap}
\end{equation}
The overlap only depends on the ratio $w_{n}/w_{0}$ between acoustic and
optical waists.

We assume for simplicity that the loss angles $\Phi _{n,p}\left[ \Omega %
\right] $\ are the same for all modes and are constant in frequency.\ As a
consequence these loss angles are simply related to the quality factor $Q$\
of the fundamental mode $\left\{ 1,0\right\} $\ of the resonator by 
\begin{equation}
\Phi _{n,p}\left[ \Omega \right] \equiv 1/Q\text{.}
\end{equation}

These equations allow to compute any parameters of the optomechanical
coupling.\ For example the non-linear phase shift $\Psi _{NL}$ is related
through equation (\ref{Eq_PsiNL2}) to the effective susceptibility at zero
frequency which is given by 
\begin{equation}
\chi _{eff}\left[ 0\right] =\sum_{n,p}\frac{\left\langle
v_{0}^{2},u_{n,p}\right\rangle ^{2}}{M_{n}\Omega _{n,p}^{2}}\text{.}
\label{Eq_ChiEff0}
\end{equation}
This sum can be numerically computed.\ In the following we consider a small
plano-convex resonator of thickness $h_{0}$ equal to 1.5\ mm and with a
curvature radius $R$ of the convex side equal to 150 mm. For these values
the acoustic waist $w_{1}$ of the fundamental mode is equal to 3.8 mm.\ It
is much larger than the optical waist $w_{0}$ which is typically of the
order of 100 $\mu $m for a cavity of 1 mm long with a curvature radius of
the input mirror equal to 1 m.\ As a consequence the overlap (\ref
{Eq_Overlap}) slowly decreases with the mode indexes $\left\{ n,p\right\} $
and it is necessary to sum over more than $10^{6}$ modes to obtain a
precision better than $10^{-3}$.

\subsection{Optimum squeezing}

We determine now the optimum squeezing $S_{opt}\left[ \Omega \right] $ of
the reflected field, that is the minimum value of the spectrum $S_{\theta
}^{out}\left[ \Omega \right] $ obtained at every frequency $\Omega $ by
scanning the quadrature $\theta $. We assume that the mechanical resonator
is made of silica (density $\rho =2200$ kg/m$^{3}$, longitudinal sound
velocity $c_{l}=5960$ m/s). Its fundamental resonance frequency $\Omega
_{M}/2\pi $ is then equal to 2 MHz and we take a quality factor $Q$ of $%
10^{6}$. The cavity is characterized by its damping $\gamma $ equal to $%
10^{-5}$ (cavity finesse ${\cal F}=3\times 10^{5}$) and its working point is
defined by the mean detuning $\overline{\Psi }$ chosen equal to $-0.2\gamma $%
. We also assume that the cavity bandwidth $\Omega _{cav}=\gamma /\tau $ is
equal to $\Omega _{M}$ and that the optical wavelength $\lambda $ is equal
to 800 nm.

To determine the non-linear phase shift $\Psi _{NL}$, we compute the
effective susceptibility $\chi _{eff}\left[ 0\right] $ at zero frequency
from equation (\ref{Eq_ChiEff0}).\ We obtain $\chi _{eff}\left[ 0\right]
=1.4\times 10^{-8}$ m/N for an optical waist $w_{0}$ of 200 $\mu $m.\ From
equations (\ref{Eq_PsiNL2}) and (\ref{Eq_I}) one then gets a non-linear
phase shift $\Psi _{NL}$ equal to $0.28\gamma $ for an incident power $%
P^{in} $ of 10 mW. Note that these values correspond to a positive slope $%
\sigma $ of the bistability curve, equal to $0.93\gamma^{2}$ (eq. \ref
{Eq_Sigma}). The working point is thus on a stable branch of the bistability
curve.

\begin{figure}
\centerline{\psfig{figure=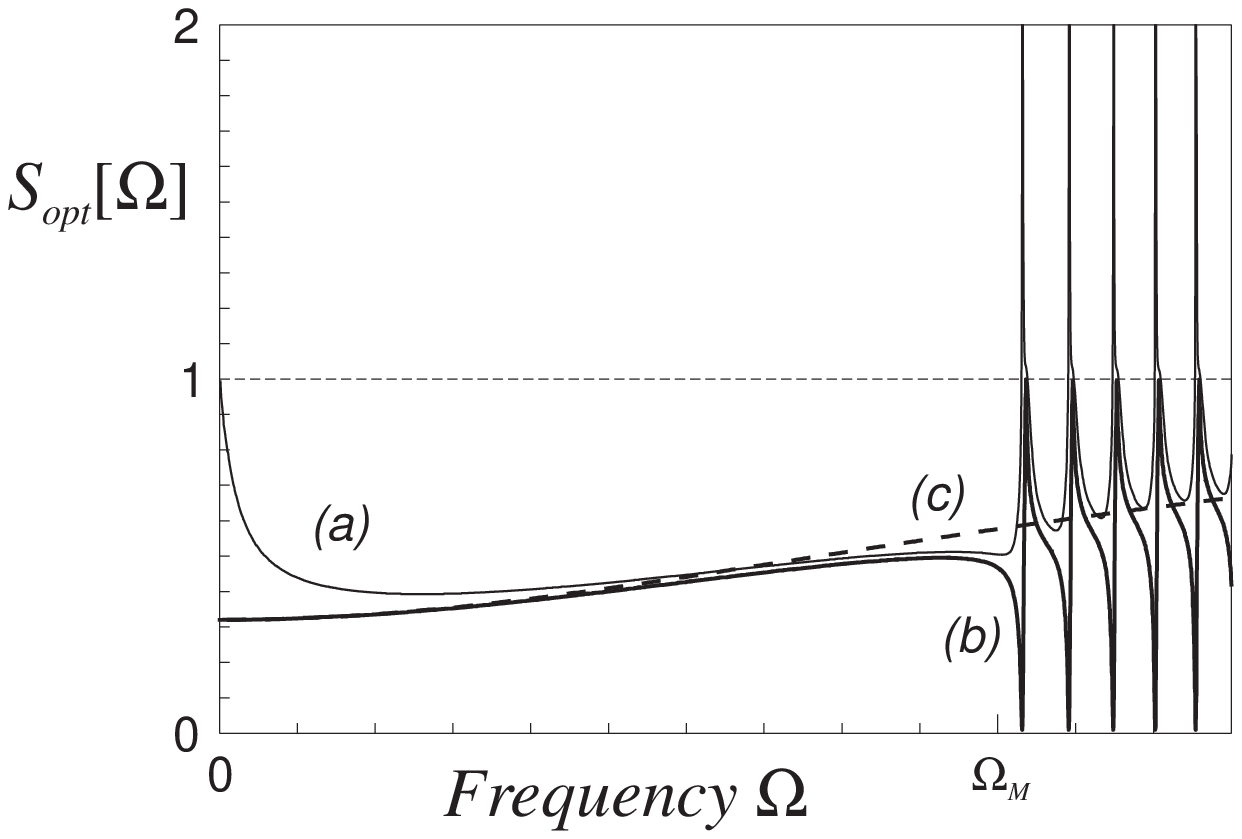,width=8 cm}}
\caption{Optimum noise spectrum $S_{opt}\left[ \Omega \right] $ of the reflected field
as a function of frequency. Curves (a) and (b) are obtained at $T=4K$ and at zero
temperature respectively. Dashed curve (c) corresponds to the pure Kerr effect for the
same parameters ($\overline{\Psi }=-0.2\gamma $, $\Psi _{NL}=0.28\gamma $)}
\label{Fig_skerr}
\end{figure} 

Figure \ref{Fig_skerr} shows the optimum squeezing $S_{opt}\left[ \Omega %
\right] $ as a function of frequency. Curve (a) is obtained at a temperature
of 4 K.\ The noise spectrum exhibits a strong reduction over a wide
frequency range from zero up to the first resonance frequency of the
resonator. For higher frequencies the spectrum shows an excess noise at
every mechanical resonance frequencies.

It is instructive to compare this spectrum to the ones obtained at zero
temperature (curve b) and for a pure Kerr effect with the same parameters $%
\Psi _{NL}$ and $\overline{\Psi }$ (curve c). At low frequencies ($\Omega
<\Omega _{M}$) the noise spectrum at zero temperature (curve b) is similar
to the one obtained with a Kerr effect (curve c). The whole effect of the
mechanical motion can be interpreted as a non-linear phase shift $\Psi _{NL}$
for the light.\ At finite temperatures (curve a), the Brownian motion of the
mirror slightly increases the noise.\ Note that this effect depends on the
dissipation mechanisms in the resonator.\ In particular the large increase
at zero frequency is due to the choice of a constant loss angle.\ One would
obtain less thermal noise in the framework of a Navier-Stockes model for
which the loss angle is a linear function of frequency.

For higher frequencies $\left( \Omega >\Omega _{M}\right) $ the dynamics of
the mechanical resonator plays an important role. At zero temperature (curve
b) we observe a series of dispersion shaped resonances centered on every
mechanical resonance $\Omega _{n,p}$, with an important noise reduction for
frequencies slightly below each resonance. Thermal noise however masks this
behavior since the Brownian motion is concentrated around the mechanical
resonances and increases the noise (curve a).

To summarize this discussion, it appears that the most interesting frequency
domain for quantum noise reduction is the low frequency domain ($\Omega
<\Omega _{M}$).\ For sufficiently low temperature the noise behavior is
similar to the one obtained with a pure Kerr effect and it mainly depends on
the non-linear phase shift $\Psi _{NL}$. More precisely the frequency
dependence of the noise spectrum roughly corresponds to a low-pass filtering
due to the cavity bandwidth and the amplitude of noise reduction becomes
important near the bistability turning points\cite{Reynaud89}.\ To reach
these points the non-linear phase shift $\Psi _{NL}$ must be large enough,
that is of the order of $\gamma $.

\subsection{Effective mass}

\begin{figure}
\centerline{\psfig{figure=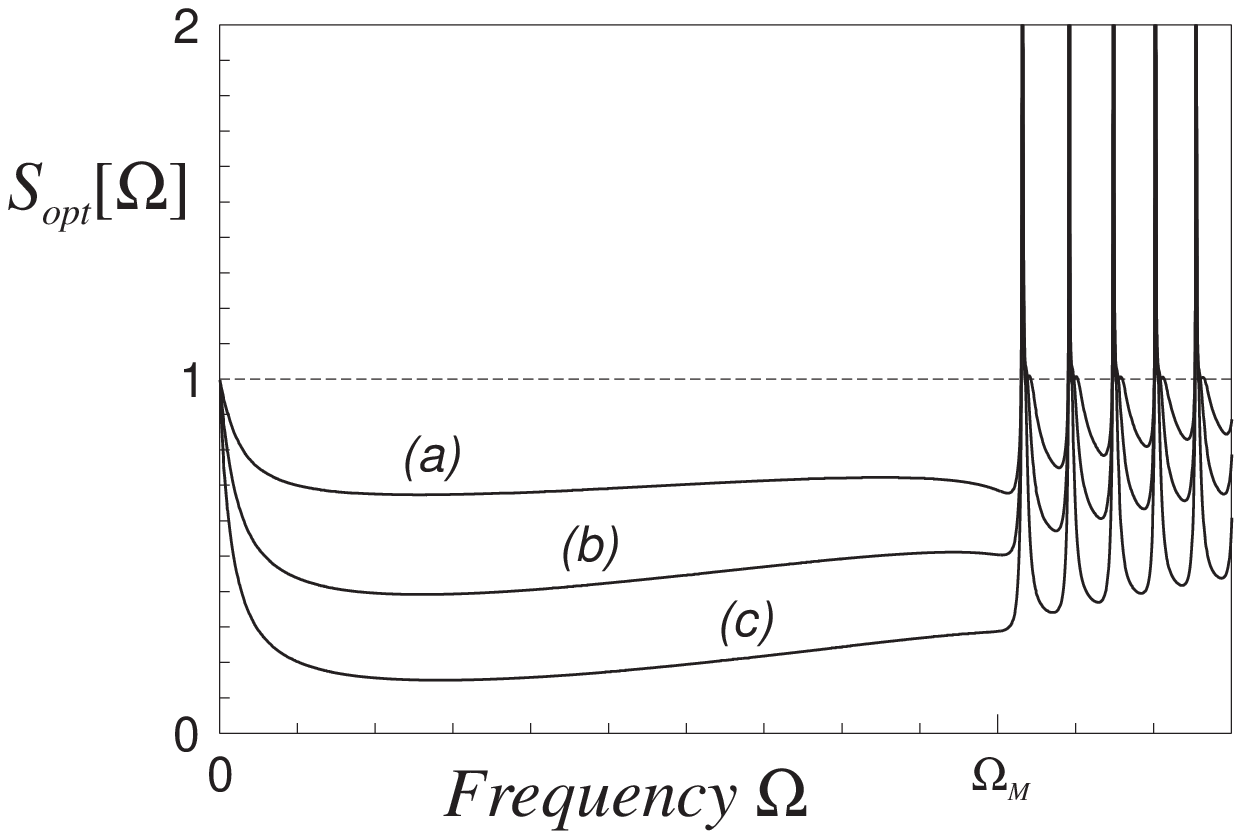,width=8cm}}
\caption{Optimum noise spectra $S_{opt}\left[ \Omega \right] $ of the reflected field
for different optical waists $w_{0}$. Curves (a) to (c) correspond respectively to 400,
200 and 100 $\mu$m}
\label{Fig_sw0}
\end{figure}

The non-linear phase shift depends on the mechanical and optical properties
of the system. In particular it is related to the spatial overlap between
the various acoustic modes and the light (eq.\ \ref{Eq_ChiEff0}).\ Figure 
\ref{Fig_sw0} shows the optimum squeezing $S_{opt}\left[ \Omega \right] $
for different values of the optical waist $w_{0}$ (400, 200 and 100 $\mu $%
m), the other parameters being identical to the ones of figure \ref
{Fig_skerr} ($P_{in}=10$ mW, $\gamma =10^{-5},$ $\overline{\Psi }=-0.2\gamma 
$, $T=4$ K and $Q=10^{6}$). For each curve we have computed the effective
susceptibility $\chi _{eff}\left[ 0\right] $ at zero frequency and we have
determined the non-linear phase shift $\Psi _{NL}$. Curves (a) to (c)
correspond to increasing non-linear phase shifts ($0.11\gamma $, $0.28\gamma 
$ and $0.62\gamma $, respectively) and to decreasing slopes of the
bistability curve ($0.99\gamma ^{2}$, $0.93\gamma ^{2}$ and $0.79\gamma ^{2}$%
, respectively).

One observes a drastic increase of the noise reduction when $w_{0}$ decreases%
. This result can be interpreted as a reduction of the effective mass
associated with the optomechanical coupling.\ Quantum effects observed in
figures \ref{Fig_skerr} and \ref{Fig_sw0} in the low frequency domain are
actually similar to the ones obtained with a harmonically suspended mirror.\
The effective susceptibility of the resonator is thus equivalent at low
frequency to the susceptibility of a single harmonic oscillator with a
resonance frequency $\Omega _{M}$.\ The mass of this equivalent pendulum is
however different from the total mass of the resonator since it depends on
the spatial matching with the light.\ This effective mass $M_{eff}$ can be
deduced from a comparison between the effective susceptibility $\chi _{eff}%
\left[ 0\right] $ at zero frequency and the one of a harmonic oscillator
(eqs.\ \ref{Eq_ChiOH} and \ref{Eq_ChiEff0}) 
\begin{equation}
\frac{1}{M_{eff}\Omega _{M}^{2}}=\sum_{n,p}\frac{\left\langle
v_{0}^{2},u_{n,p}\right\rangle ^{2}}{M_{n,p}\Omega _{n,p}^{2}}\text{.}
\label{Eq_Meff}
\end{equation}
The effective mass is related to the masses of all acoustic modes coupled to
the light. The number of acoustic modes contributing to the sum strongly
depends on the light waist.\ The typical variation length of the Laguerre
polynomial $L_{p}\left( 2r^{2}/w_{n}^{2}\right) $ in the spatial structure
of the acoustic mode $\left\{ n,p\right\} $ is $w_{n}/\sqrt{p}$. The overlap 
$\left\langle v_{0}^{2},u_{n,p}\right\rangle $ is thus equal to $1$ as long
as $w_{0}$ is much smaller than $w_{n}/\sqrt{p}$ (eq.\ \ref{Eq_Overlap}).
For a given value of $n$, the number of transverse modes which contribute to
the effective mass is proportional to $w_{1}^{2}/nw_{0}^{2}$ and it
increases when $w_{0}$ decreases. As a consequence the mass becomes smaller
and quantum effects of radiation pressure become larger.

\begin{figure}
\centerline{\psfig{figure=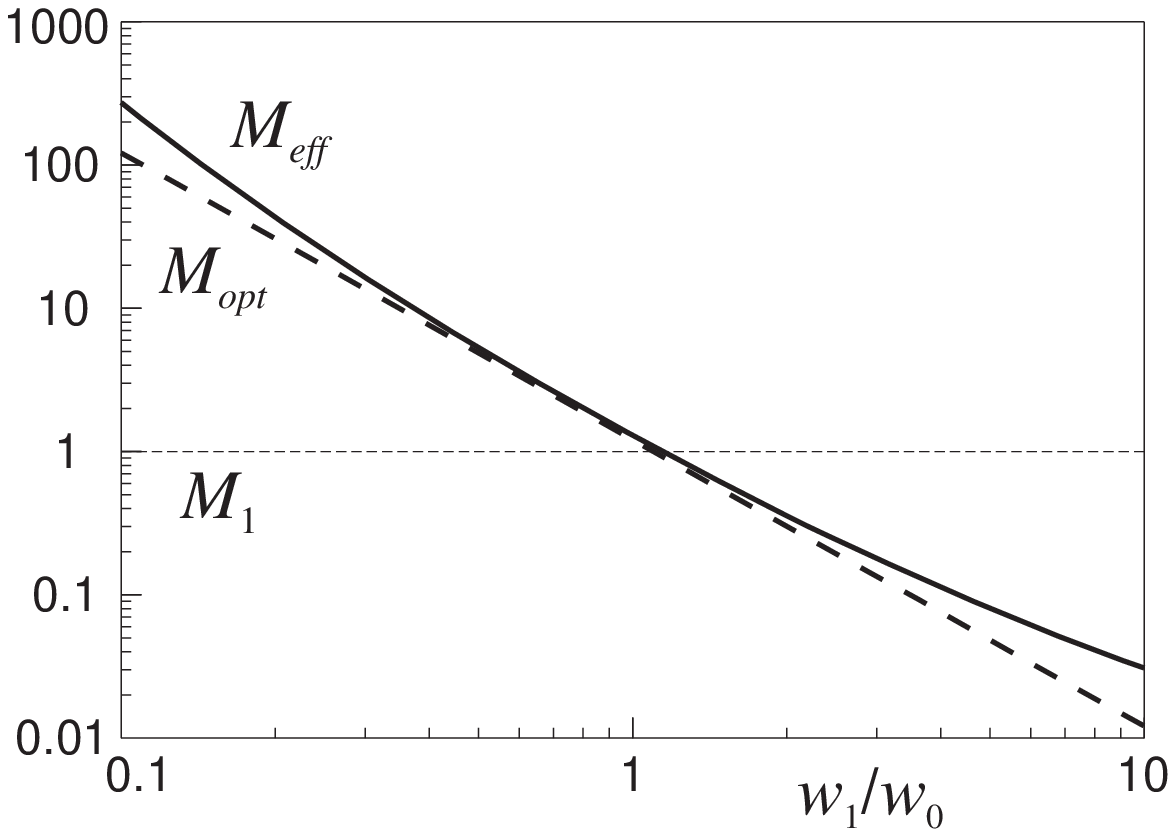,width=8 cm}}
\caption{Effective mass $M_{eff}$ of the resonator as a function of the ratio between
fundamental acoustic and optical waists. Dashed curve represents the optical mass $M_{opt}$
related to the volume of the resonator illuminated by the light. Vertical scale is
normalized to the mass $M_{1}$ of the fundamental acoustic mode (37 mg)}
\label{Fig_mass}
\end{figure}

From equations (\ref{Eq_Onp}) to (\ref{Eq_Overlap}) one can express the
effective mass as a function of the waists $w_{1}$, $w_{0}$ and of the mass $%
M_{1}$ of the fundamental mode 
\begin{eqnarray}
\frac{M_{1}}{M_{eff}} &=&\sum_{n,p}\frac{4w_{1}^{4}}{\left(
2w_{1}^{2}+nw_{0}^{2}\right) ^{2}}\left( \frac{2w_{1}^{2}-nw_{0}^{2}}{%
2w_{1}^{2}+nw_{0}^{2}}\right) ^{2p}\times  \nonumber \\
&&\times \frac{1}{n+\frac{2}{\pi }\sqrt{\frac{h_{0}}{R}}\left( 2p+1\right) }%
\text{.}  \label{Eq_M1Meff}
\end{eqnarray}
Solid line in figure \ref{Fig_mass} shows the variation of the effective
mass as a function of the ratio $w_{1}/w_{0}$ for a resonator of thickness $%
h_{0}=1.5$ mm and of curvature radius $R=150$ mm. It clearly appears that
this mass decreases with the optical waist and very small values can be
reached.\ The mass $M_{1}$ of the fundamental mode is equal to 37 mg (eq.\ 
\ref{Eq_Mnp}) and one gets an effective mass of 1 mg for an optical waist
equal to $w_{1}/10$ (380 $\mu $m) and a mass of 0.2 mg for a waist of 100 $%
\mu $m. Such small values would be very difficult to obtain with a
harmonically suspended mirror for which the mass associated with the global
motion is the total mass of the mirror.

One can get a simple physical insight into the effective mass in the case of
a resonator thickness much smaller than the curvature radius of the convex
side ($h_{0}\ll R$). We can then assume that the transverse modes of the
resonator are degenerate, that is the resonance frequencies $\Omega _{n,p}$
are independent of $p$. The sum over $p$ in equation (\ref{Eq_M1Meff}) is a
simple geometric sum and one gets an estimate $M_{opt}$ of the effective
mass given by 
\begin{equation}
\frac{M_{1}}{M_{opt}}=\frac{w_{1}^{2}}{2w_{0}^{2}}\sum_{n}\frac{1}{n^{2}}=%
\frac{\pi ^{2}}{12}\left( \frac{w_{1}}{w_{0}}\right) ^{2}\text{.}
\end{equation}
Using the expression of $M_{1}$ (eq. \ref{Eq_Mnp}) we finally obtain 
\begin{equation}
M_{opt}=\frac{12}{\pi ^{2}}\left( \frac{\pi }{4}\rho h_{0}w_{0}^{2}\right) 
\text{.}
\end{equation}
The term in brackets corresponds to the mass of the part of the resonator
illuminated by the light beam. This optical mass is a good approximation of
the effective mass $M_{eff}$ as shown by the dashed curve in figure \ref
{Fig_mass}.\ We have thus shown that the effect of the resonator motion on
the light is equivalent to the one of a harmonically suspended mirror of
mechanical resonance frequency $\Omega _{M}$ and of mass related to the
light spot size.\ This mass can of course become very small for a small
optical waist.

\section{Conclusion}

We have studied the quantum effects due to radiation pressure in a
high-finesse cavity with a mirror coated on a mechanical resonator.\ We have
shown that the optomechanical coupling{\it \ }between the gaussian laser
beam and the acoustic modes of the resonator leads to a non-linear phase
shift for the light.\ This phase shift is related to the intracavity
intensity through an effective susceptibility which takes into account all
the acoustic modes and their coupling to the light. This susceptibility also
describes the effect on the light of the Brownian motion of the mirror.

We have studied the quantum noise reduction of the field reflected by the
cavity.\ This quantum effect mainly depends on the behavior of the
non-linear phase shift at low frequency.\ In this frequency domain the
mechanical response of the resonator can be approximated to a harmonic
response. The effect of optomechanical coupling is then equivalent to the
one obtained with a harmonically suspended mirror of resonance frequency
equal to the fundamental resonance frequency of the resonator.\ The mass of
this equivalent pendulum is however smaller than the total mass of the
mirror, thus enhancing the optomechanical coupling. We have shown that for a
plano-convex resonator this effective mass is of the order of the optical
mass which corresponds to the volume of the resonator illuminated by the
light.\ The effective mass can be two or three orders of magnitude smaller
than the total mass of the mirror and large quantum effects are obtained for
a reasonable input power when the optical waist is small enough.

This drastic decrease of the effective mass seems to be a specific behavior
of the plano-convex geometry.\ The effective susceptibility of a cylindrical
mirror has already been determined at low frequency for the study of thermal
effects in gravitational-wave detectors\cite{Bondu95,Gillepsie95}.\ The
reduction of the effective mass does not exceed a factor 10 below the total
mass of the mirror.\ The plano-convex geometry thus allows to get
simultaneously a high resonance frequency and a very small mass so that\
large quantum noise reduction can be obtained over a wide frequency range.
One can of course obtain significant non-linear phase shifts with a
cylindrical mirror but the fundamental resonance frequency is very low.

Finally, we have only studied in this paper the quantum noise reduction of
the field reflected by the cavity.\ Similar results would however be
obtained for other quantum effects of radiation pressure such as the
possibility of quantum nondemolition measurement of light intensity\cite
{Heidmann97} or the quantum limit in interferometric measurements.\ For
example it has been shown that the quantum limit for measurement sensitivity
is proportional to the susceptibility characterizing the mirror motion\cite
{Jaekel90}.\ The quantum limit induced by internal acoustic modes is thus
proportional to the effective susceptibility studied in this paper.\ In
particular the sensitivity is reduced if mirrors have small effective
masses.\bigskip

{\bf Acknowledgements}

We gratefully acknowledge F.\ Bondu for the program {\it CYPRES} used to
compute the effective susceptibility of a cylindrical mirror.\ Y. Hadjar\
acknowledges a fellowship from the {\it Association Louis de Broglie d'Aide
\`{a} la Recherche}.\bigskip

\appendix

\section{Propagation equation for a resonator submitted to radiation pressure%
}

In this appendix we derive the evolution equation of the acoustic modes for
a resonator submitted to a radiation pressure force.\ This derivation is
actually similar to the phonon decomposition in presence of an external
force.

The evolution equation of any displacement $\overrightarrow{u}\left( 
\overrightarrow{r},t\right) $ can be deduced from the Lagrangian $L=T-U$
where the kinetic energy $T$ and the potential energy $U$ are equal to\cite
{Landau58b} 
\begin{mathletters}
\label{EqsA_Energy}
\begin{eqnarray}
T &=&\int_{V}\frac{1}{2}\rho \left| \frac{\partial \overrightarrow{u}}{%
\partial t}\right| ^{2}d^{3}r\text{,}  \label{EqA_EnergyT} \\
U &=&\sum_{ij}\int_{V}\frac{1}{2}\sigma _{ij}u_{ij}d^{3}r\text{,}
\label{EqA_EnergyU}
\end{eqnarray}
where $\rho $ is the density of the resonator. The strain tensor $u_{ij}$ is
given by 
\end{mathletters}
\begin{equation}
u_{ij}=\frac{1}{2}\left( \frac{\partial u_{j}}{\partial x_{i}}+\frac{%
\partial u_{i}}{\partial x_{j}}\right) \text{,}
\end{equation}
where $u_{i}$ ($i=1$, 2, 3) are the cartesian components of the displacement 
$\overrightarrow{u}\left( \overrightarrow{r},t\right) $ at point $%
\overrightarrow{r}$ of cartesian coordinates $x_{i}$. The stress tensor $%
\sigma _{ij}$ is related to the strain tensor $u_{ij}$ by Hooke's law 
\begin{equation}
\sigma _{ij}=2\mu u_{ij}+\lambda \overrightarrow{\nabla }.\overrightarrow{u}%
\delta _{ij}\text{,}
\end{equation}
where $\lambda $ and $\mu $ are the Lam\'{e} constants of the resonator. We
deduce from the Lagrange equations the propagation equation of $%
\overrightarrow{u}\left( \overrightarrow{r},t\right) $ in the resonator 
\begin{equation}
\rho \frac{\partial ^{2}\overrightarrow{u}}{\partial t^{2}}=\mu \Delta 
\overrightarrow{u}+\left( \lambda +\mu \right) \overrightarrow{\nabla }%
\left( \overrightarrow{\nabla }.\overrightarrow{u}\right) \text{.}
\label{EqA_Propag}
\end{equation}
For a free resonator the constraint tensor $\sigma _{ij}$ at every point $%
\overrightarrow{r}$ of the surface must satisfy the following condition for
any cartesian index $i$ 
\begin{equation}
\sum_{j}\sigma _{ij}\left( \overrightarrow{r},t\right) n_{j}=0\text{,}
\label{EqA_Constraint}
\end{equation}
where $\overrightarrow{n}$ is the normal vector at point $\overrightarrow{r}$%
.

Acoustic modes are defined as monochromatic solutions $\overrightarrow{u}%
_{n}\left( \overrightarrow{r}\right) e^{-i\Omega _{n}t}$ of equations (\ref
{EqA_Propag}) and (\ref{EqA_Constraint}).\ Each acoustic mode can be
decomposed into longitudinal and transverse components noted $%
\overrightarrow{u}_{n}^{l}$ and $\overrightarrow{u}_{n}^{t}$ respectively
and defined by 
\begin{equation}
\overrightarrow{\nabla }\times \overrightarrow{u}_{n}^{l}=0\text{, }%
\overrightarrow{\nabla }.\overrightarrow{u}_{n}^{t}=0\text{.}
\end{equation}
These components obey the following propagation equations 
\begin{mathletters}
\label{EqsA_Modes}
\begin{eqnarray}
\Delta \overrightarrow{u}_{n}^{l}+\frac{\Omega _{n}^{2}}{c_{l}^{2}}%
\overrightarrow{u}_{n}^{l} &=&0\text{,} \\
\Delta \overrightarrow{u}_{n}^{t}+\frac{\Omega _{n}^{2}}{c_{t}^{2}}%
\overrightarrow{u}_{n}^{t} &=&0\text{,}
\end{eqnarray}
where $c_{l}$ and $c_{t}$ are the longitudinal and transverse sound
velocities (eq. \ref{Eqs_Velocity}).

Since the set $\left\{ \overrightarrow{u}_{n}\left( \overrightarrow{r}%
\right) \right\} $ of acoustic modes forms a basis for the solutions of
equations (\ref{EqA_Propag}) and (\ref{EqA_Constraint}), any displacement $%
\overrightarrow{u}\left( \overrightarrow{r},t\right) $ can be decomposed on
these modes with time-dependent amplitudes $a_{n}\left( t\right) $ (eq.\ \ref
{Eq_Decomp}).\ We now determine the total energy $E=T+U$ associated with
this displacement.\ From equation (\ref{EqA_EnergyT}) the kinetic energy $T$
is equal to 
\end{mathletters}
\begin{equation}
T=\sum_{n}\frac{1}{2}M_{n}\left( \frac{da_{n}}{dt}\right) ^{2}\text{,}
\end{equation}
where $M_{n}$ is the mass of mode $u_{n}$ (eq.\ \ref{Eq_Mn}). The potential
energy $U$ is given by equation (\ref{EqA_EnergyU}) and can be written as 
\begin{equation}
U=\sum_{i,j}\frac{1}{2}\int_{V}d^{3}r\left[ \frac{\partial \left( \sigma
_{ij}u_{j}\right) }{\partial x_{i}}-\frac{\partial \sigma _{ij}}{\partial
x_{i}}u_{j}\right] \text{.}
\end{equation}
The first term can be transformed into a surface integral which is equal to
zero for a displacement satisfying the boundary condition (\ref
{EqA_Constraint}). Using the decomposition of $\overrightarrow{u}_{n}$ into
longitudinal and transverse components and the propagation equations (\ref
{EqsA_Modes}) one then gets 
\begin{equation}
U=\sum_{n}\frac{1}{2}M_{n}\Omega _{n}^{2}\left[ a_{n}\left( t\right) \right]
^{2}\text{.}
\end{equation}
The total energy $E$ is finally equal to 
\begin{equation}
E=\sum_{n}\left\{ \frac{1}{2}M_{n}\left( \frac{da_{n}}{dt}\right) ^{2}+\frac{%
1}{2}M_{n}\Omega _{n}^{2}\left[ a_{n}\left( t\right) \right] ^{2}\right\} 
\text{,}
\end{equation}
and appears as the sum of the energies of free harmonic oscillators.

Up to now we have assumed that the resonator is free from any external
constraint.\ In presence of radiation pressure there is an additional
contribution to the energy corresponding to the work of internal constraints
opposed to the external force 
\begin{equation}
W=-\int_{z=0}d^{2}r\overrightarrow{F}_{rad}\left( \overrightarrow{r}%
,t\right) .\overrightarrow{u}\left( \overrightarrow{r},t\right) \text{.}
\end{equation}
The total energy $E$ is then equal to 
\begin{eqnarray}
E &=&\sum_{n}\left\{ \frac{1}{2}M_{n}\left( \frac{da_{n}}{dt}\right) ^{2}+%
\frac{1}{2}M_{n}\Omega _{n}^{2}\left[ a_{n}\left( t\right) \right]
^{2}\right.  \nonumber \\
&&\left. -\left\langle \overrightarrow{F}_{rad},\overrightarrow{u}%
_{n}\right\rangle a_{n}\left( t\right) \right\} \text{,}
\end{eqnarray}
where $\left\langle \overrightarrow{F}_{rad},\overrightarrow{u}%
_{n}\right\rangle $ is the overlap integral in the $z=0$ plane of the scalar
product between the radiation pressure and the acoustic mode $u_{n}$ (eq.\ 
\ref{Eq_Scalar}).\ We have thus shown that the total energy\ is the sum over
all modes of the energies of forced harmonic oscillators. The evolution
equation (\ref{Eq_Motion_an}) for each mode amplitude $a_{n}$ is then
deduced from Hamilton's equations.

\end{document}